\documentclass[12pt]{article}
\usepackage{amsmath,amssymb,amsthm,color}
\usepackage{graphicx}
\usepackage{cite}
\usepackage{epsfig}
\renewcommand{\baselinestretch}{1.66}

\begin{document}

\title {Coulomb scattering rates of excited states in germanene \\ }

\author{
\small Po-Hsin Shih$^{a}$, Chih-Wei Chiu$^{b}$, Jhao-Ying Wu$^{c,*}$, Thi-Nga Do$^{b}$, Ming-Fa Lin$^{a,*}$\\
\small  $^a$Department of Physics, National Cheng Kung University, Tainan, Taiwan 701\\
\small  $^b$Department of Physics, National Kaohsiung Normal University, Kaohsiung, 824 Taiwan\\
\small  $^c$Center of General Studies, National Kaohsiung Marine University, Kaohsiung, 811 Taiwan\\
 }

\renewcommand{\baselinestretch}{1.66}
\maketitle

\renewcommand{\baselinestretch}{1.66}

\begin{abstract}

The excited conduction electrons, conduction holes and valence holes in monolayer germanene exhibit the feature-rich Coulomb decay rates.
The dexcitation processes are studied using the Matsubara's screened exchange energy.
They might utilize the intraband single-particle excitations (SPEs), the interband SPEs, and three kinds of plasmon modes, depending on the quasiparticle states and the Fermi energies.
The low-lying valence holes can decay by the undamped acoustic plasmon, so that they present very fast Coulomb deexcitations, the non-monotonous energy dependence and the anisotropic behavior.
However, the low-energy conduction holes and electrons behave as 2D electron gas.
The high-energy conduction states and the deep-energy valence ones are similar in the available deexcitation channels and the dependence of decay rate on wave vector ${\bf k}$.

\end {abstract}
\renewcommand{\baselinestretch}{1.66}
\maketitle

\newpage
A lot of two-dimensional (2D) materials have been successfully synthesized since the first discovery of graphene in 2004 using the mechanical exfoliation of Bernal graphite\cite{SCI:306-666}.
They are very suitable for exploring the diverse physical, chemical, and material properties.
Specifically, the 2D IV-group systems possess the high-symmetry honeycomb lattice and the nano-scaled thickness, in which few-layer graphenes have been verified to exhibit the rich and unique properties, such as the massless/massive fermions \cite{Na:438-197,PRL:102-808,RMP:81-109,PR:71-622}, the quantized Landau levels \cite{PRL:98-197403,PRL:110-076801,PRB:77-426,PCCP:15-26008}, the magneto-optical selection rules \cite{NL:7-2711,PRB:85-403,PRB:77-313,arXiv:1603}, and the quantum Hall effects\cite{NATURE:438-201,Sci:340-167,PRL:95-801,arXiv:1704}.
Recently, few-layer germanene, silicene and tinene are, respectively, grown on [Pt(111), Au(111) $\&$ Al(111)] surfaces\cite{AM:26-820,NJP:16-2,TDM:4-005,NL:15-510},  [Ag(111), Ir(111) $\&$ Zr$\mathrm{Bi_2}$] surfaces\cite{PRL:108-501,NL:13-685,PRL:108-5501}, and $\mathrm{Bi_2Te_3}$(111) surface\cite{NM:14-1020}.
Such systems possess the buckled structures and the significant spin-orbital couplings (SOC’s), leading to the dramatic changes in the essential properties.
They are expected to present the unusual Coulomb excitations/deexcitations arising from many-particle electron-electron interactions.
The Coulomb scattering rates of the excited states in monolayer germanene is chosen for a model study in this work, especially for their relations with the single-particle and collective electronic excitations.\\

For germanene, silicene and graphene, the low-lying electronic structures mainly arise from the outmost $\mathrm{p_z}$ orbitals\cite{RMP:81-109,PRB:84-430}.
 The Dirac-cone structures, being created by the hexagonal symmetry, might be  separated or gapless as a result of the significant/negligible SOC’s.
The former two are predicted to be narrow-gap semiconductors (${E_g\sim}$ 93 meV for Ge $\&$ $\sim$7.9 meV for Si), reflecting the strength of SOC\cite{PRB:84-430}.
However, graphene has linear valence and conduction bands intersecting at the Dirac point in the absence of SOC.
The predicted band structures could be verified from the angle-resolved photoemission spectroscopy (ARPES) measurements, as done for few-layer germanene grown on Au(111) surface\cite{TDM:4-005}.
The experimental observations show that the multiple Dirac-like energy dispersions might be caused by the folding of germanene's Dirac cones.
The high-resolution ARPES measurements could also provide the full information on the energy widths of the excited states\cite{TDM:4-005,NP:3-36,PRL:98-802}.\\

The electron-electron Coulomb interactions are one of the main-stream topics in condensed-matter systems\cite{PRB:81-406,NANO:7-620,CM:25-34,PRB:75-418,PRB:89-410,SR:7-600,PRB:74-406}.
They can create the many-particle electronic excitations and thus have strong effects on the energies and lifetimes of quasi-particle states.
The previous calculations predict that monolayer germanene exhibits  the diverse momentum- and frequency-dependent phase diagrams\cite{PRB:89-410,SR:7-600}.
The feature-rich Coulomb excitations, as shown Figs. 1(a) and 1(b) at ${E_F=0.2}$ eV,  cover the anisotropic excitation spectra, the intraband single-particle excitations (SPEs), the interband SPEs, the strong acoustic plasmon at small momenta (q's), the second kind of plasmon (the undamped mode at large q's by the blue arrow in Fig. 1(a)), and the third kind of plasmon accompanied with the intraband Landau damping (the purple arrow in Fig. 1(b)).
They might become the effective deexcitation channels of the excited electrons/holes, being closely related to the wave vectors, valence/conduction states and Fermi energies. This is worthy of a systematic investigation.\\

The Matsubara's screened exchange energy is used to calculate the Coulomb scattering rates of the excited states in monolayer  germanene, in which the deexcitation channels are evaluated from the random-phase approximation (RPA).
The decay processes and their dependence on the wave vector, valence/conduction states, and Fermi energy are explored in detail.
This work shows that the intraband SPEs, the interband SPEs, and the diverse plasmon modes play critical roles in determining the deexcitation behaviors.
The feature-rich Coulomb decay rates cover the oscillatory energy dependence, the non-equivalent valence and conduction Dirac points, the strong anisotropy, and the similarity with 2D electron gas for the  low-energy conduction electrons and valence holes.
The predicted Coulomb decay rates could be verified from the ARPES measurements on the energy widths of quasiparticle states\cite{TDM:4-005,NP:3-36,PRL:102-803}.\\

Monolayer germanene has a buckled hexagonal lattice with the Ge-Ge bond length of $b=2.32$ $\AA$.
There are two equivalent sublattices of A and B, being separated by a distance of $l=0.66$ $\AA$ (details in \cite{SR:7-600}).
The low-lying electronic structure is dominated by $4p_z$ orbitals.
The Hamiltonian, which is built from the sub-space spanned by the four spin-dependent tight-binding functions, is expressed as

\begin{equation}
H=-t\displaystyle\sum_{\langle i,j \rangle, \alpha}c^{\dagger}_{i\alpha}c_{j\alpha}+\mathrm{i}\frac{\lambda_{so}}{3\sqrt{3}}\displaystyle\sum_{\langle \langle i,j \rangle \rangle, \alpha, \beta}\nu_{ij}c^{\dagger}_{i\alpha}\sigma^{z}_{\alpha\beta}c_{j\beta}- \mathrm{i}\frac{2}{3}\lambda_{R}\displaystyle\sum_{\langle \langle i,j \rangle \rangle, \alpha, \beta}\mu_{ij}c^{\dagger}_{i\alpha}(\vec{\sigma}\times\hat{d}_{ij})c_{j\beta}.
\end{equation}

The first term, in which the summation takes over all pairs $\langle i,j \rangle$  of the nearest-neighboring lattice sites, is the kinetic energy with the hopping integral of $t$ = 0.86 eV \cite{PRB:84-430}.
$c^{\dagger}_{i\alpha}$ $( c_{j\alpha})$ can create (annihilate) an electron with spin polarization $\alpha$ $(\beta)$ at the $i$-th ($j$-th) site.
The second term represents the effective SOC with the summation on all pairs $ \langle \langle i,j \rangle \rangle$ of the next-nearest-neighboring sites and $\lambda _{SO}$ = 46.3 meV.
 $\vec {\sigma} = (\sigma_x$, $\sigma_y$, $\sigma_z) $ is the Pauli spin matrix.
$\nu_{i,j} = ( \vec{d_i} \times \vec{d_j} )/ |\vec{d_i} \times \vec{d_j} |$, where $\nu_{i,j}=+1$ and $-1$, respectively, correspond to the  anti-clockwise and clockwise cases from the cross product of the two nearest-neighboring bonding vectors $\vec{d_i}$ and $\vec{d_j}$.
The third term denotes the Rashba SOC with $\lambda_R = 10.7$ meV, $u_{i,j} = + 1(-1)$ for the A (B) lattice sites, and $\hat{d}_{ij}$ is the unit vector connecting two sites $i$ and $j$ in the same sublattice.
State energies are characterized by ${E^{c,v}({\bf k})}$ (${E^h(\bf k\,)}$), where c and v correspond to conduction and valence states, respectively.
They remain doubly degenerate for the spin degree of freedom in the presence of SOC, in which there exists the spin-up- and spin-down-dominated configurations. \\

\begin{figure}
\center
\rotatebox{0} {\includegraphics[width=1.0\linewidth]{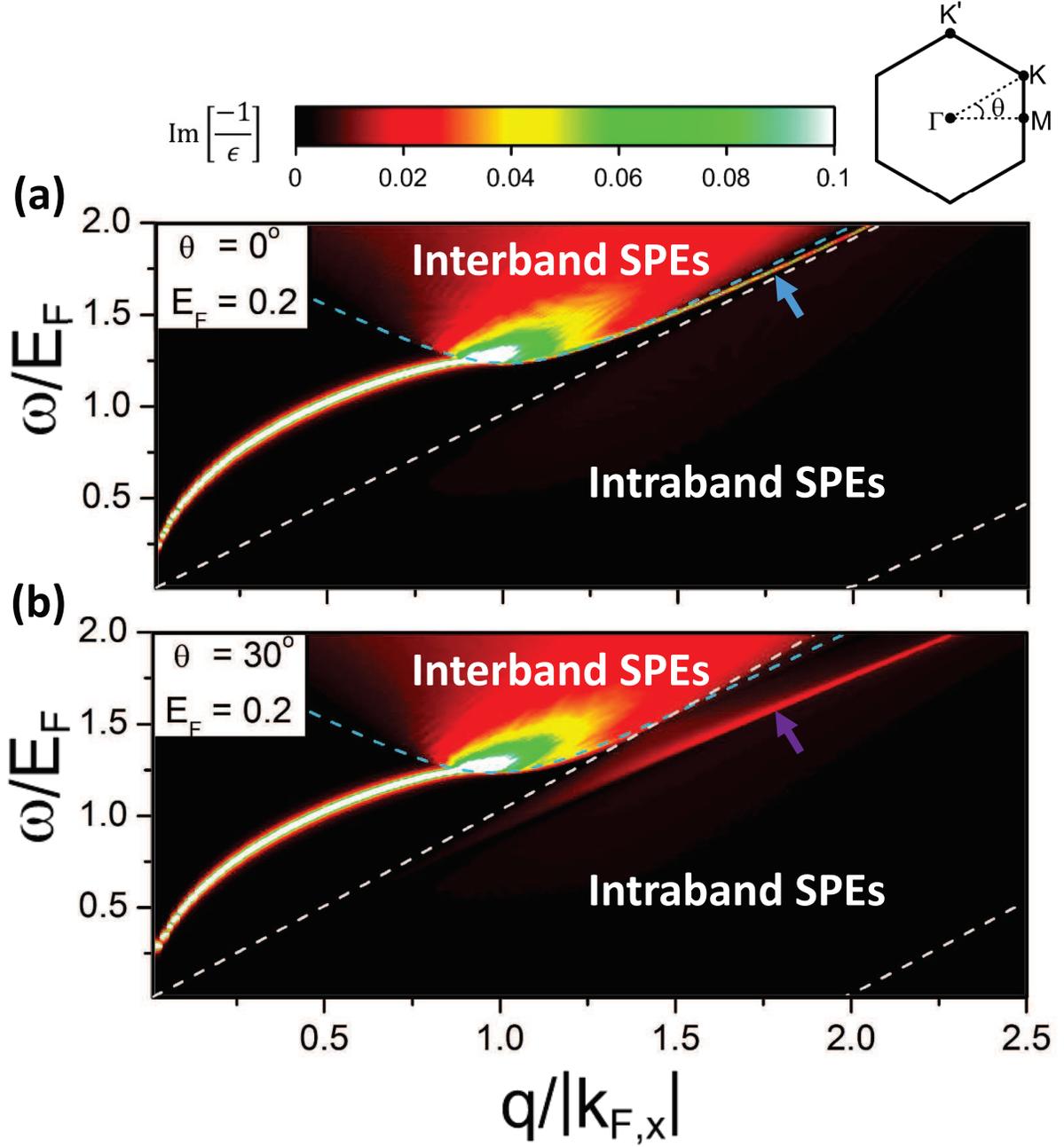}}
\caption{The momentum- and frequency-dependent excitation spectra of germanene with ${E_F=0.2}$ eV under (a) ${\theta\,=0^\circ}$ and (b) ${30^\circ}$, in which $\theta$ is the angle between the transferred momentum and $\Gamma$M. The second and third kinds of plasmon modes are, respectively, indicated by the blue and purple arrows in (a) and (b). ${k_{F,x}}$ is the Fermi momentum along $\Gamma$M.}
\label{Figure 1}
\end{figure}

The free carrier density and temperature can enrich the electronic excitations of monolayer germanene.
Under the perturbation of Coulomb interactions, the electrons can be excited from occupied states to unoccupied ones during the dynamic charge screening.
For an intrinsic germanene, only the interband SPEs, being described by the imaginary part of the dielectric function, can survive at zero temperature.
The collective excitations are revealed in the loss function as a prominent peak when the free carrier density or temperature is sufficiently high\cite{SR:7-600}.
The extrinsic germanene is predicted to exhibit three kinds of plasmon modes.
There exist intraband and interband SPEs (Figs. 1(a) and 1(b)), in which the former and the latter are, respectively, associated with the conduction and valence carriers.
The first kind of plasmon, which behaves as a 2D acoustic mode at small transferred momenta, could make much contribution to decay rates.
At large q's, it experiences the heavy interband Landau damping and then disappears.
Specifically, the second and third kinds of plasmons come to exist only under the sufficiently large momenta.
The above-mentioned single- and many-particle excitation channels are available in the Coulomb decay rates, as discussed later.\\

The incident electron beam/electromagnetic field has strong interactions with charge carriers and thus creates the excited electrons (holes) above (below) the Fermi level.
Such intermediate states could further decay by the inelastic electron-electron scatterings.
The Coulomb decay rate ($1/\tau$) is dominated by the effective interaction potential ($V^{eff}$) between two charges, in which the dynamic e-e interactions could be understood from the RPA.
By using the Matsubara Green's function\cite{Mahan},
$1/\tau$ is evaluated from the quasiparticle self-energy, the screened exchange energy
\begin{equation}
\Sigma(\mathbf{k},h,\mathrm{i}k_n)=-\frac{1}{\beta}\sum_{\mathbf{q},h',\mathrm{i}\omega_{m}}V^{eff}(\mathbf{q},\mathrm{i}\omega_m;\mathbf{k},h,h')G^{(0)}(\mathbf{k+q},h',\mathrm{i}k_{n}+\mathrm{i}\omega_{m}),
\end{equation}
where $\beta=(k_BT)^{-1}$, $\mathrm{i}k_n=\mathrm{i}(2n+1)\pi/\beta$ (fermion), $\mathrm{i}\omega_m=\mathrm{i}2m\pi/\beta$ (boson) and $G^{(0)}$ is the noninteracting Matsubara Green's function.
$V^{eff}(\mathbf{q},\mathrm{i}\omega_m;h,h',\mathbf{k})=V_q|\langle h',\mathbf{k+q}|e^{\mathrm{i}\vec{q}\cdot\vec{r}}|h,\mathbf{k}\rangle|^2/$  $[\epsilon(\mathbf{q},\mathrm{i}\omega_{m})]$ is the screened Coulomb interactions with the band-structure effect, where $V_q$ is the 2D bare Coulomb potential energy and $\epsilon(\mathbf{q},\mathrm{i}\omega_{m})$ is the dielectric function.
It should be noticed that the SOC leads to the superposition of the spin-up and the spin-down components.
However,  it does not need to deal with the spin-up- and spin-down-dependent Coulomb decay rates separately, since they make the same contribution.
That is, it is sufficient in exploring the wave-vector-, conduction/valence- and energy-dependent self-energy (Eq. (2)).
Under the analytic continuation $\mathrm{i}k_n \to E^{h}(\mathbf{k})$, the self-energy can be divided into the line part and the residue part:
\begin{equation}
\Sigma(\mathbf{k},h,E^{h}(\mathbf{k}))=\Sigma^{(line)}(\mathbf{k},h,E^{h}(\mathbf{k}))+\Sigma^{(res)}(\mathbf{k},h,E^{h}(\mathbf{k})),
\end{equation}
in which
\begin{align}
\Sigma^{(line)}(\mathbf{k},h,E^{h}(\mathbf{k}))=-\frac{1}{\beta}\sum_{\mathbf{q},h',\mathrm{i}\omega_{m}}V^{eff}(\mathbf{q},\mathrm{i}\omega_m;\mathbf{k},h,h')\\
\times G^{(0)}(\mathbf{k+q},h',E^{h}(\mathbf{k})+\mathrm{i}\omega_{m}),
\end{align}
and
\begin{align}
\Sigma^{(res)}(\mathbf{k},h,E^{h}(\mathbf{k})) =&-\frac{1}{\beta}\sum_{\mathbf{q},h',\mathrm{i}\omega_{m}}V^{eff}(\mathbf{q},\mathrm{i}\omega_m;\mathbf{k},h,h') \notag\\
&\times [G^{(0)}(\mathbf{k+q},h',\mathrm{i}k_{n}+\mathrm{i}\omega_{m}) \notag\\
&-G^{(0)}(\mathbf{k+q},h',E^{h}(\mathbf{k})+\mathrm{i}\omega_{m})].
\end{align}
The imaginary part of the residue self-energy determines the Coulomb decay rate, being defined as
\begin{align}
\mathrm{Im}\Sigma^{(res)}(\mathbf{k},h,E^{h}(\mathbf{k}))&=\frac{-1}{2\tau(\mathbf{k},h)}\notag\\
&=\sum_{q,h'}\mathrm{Im}[-V^{eff}(\mathbf{q},\omega_{de};\mathbf{k},h,h')]\notag\\
&\quad\times\{n_B(-\omega_{de})[1-n_F(E^{h'}(\mathbf{k+q}))]-n_B(\omega_{de})[n_F(E^{h'}(\mathbf{k+q}))]\}\notag\\
&=\frac{-1}{2\tau_e(\mathbf{k},h)}+\frac{-1}{2\tau_h(\mathbf{k},h)}.
\end{align}
$\omega_{de}=E^{h}(\mathbf{k})-E^{h'}(\mathbf{k+q})$ is the deexcitation energy.
$n_B$ and $n_F$ are the Bose and Fermi distribution functions, respectively.
Equation (7) indicates that an initial state of $(\mathbf{k},h)$ can be deexcited to all the available $(\mathbf{k+q},h')$ states, under the Pauli exclusion principle and the conservation laws of energy and momentum.
The excited states above or below the Fermi level are, respectively, related to the electron and hole decay rates (the first and second terms in Eq. (7)).
By detailed calculations, the zero-temperature Coulomb decay rate of the excited electrons and holes are
\begin{align}
\frac{1}{\tau_e(\mathbf{k},h)}+\frac{1}{\tau_h(\mathbf{k},h)}&=-2\sum_{q,h'}\mathrm{Im}[-V^{eff}(\mathbf{q},\omega_{de};\mathbf{k},h,h')]\notag\\
&\quad\times[-\Theta(\omega_{de})\Theta(E^{h'}(\mathbf{k+q})-E_F)+\Theta(-\omega_{de})\Theta(E_F-E^{h'}(\mathbf{k+q}))].
\end{align}
where $E_F$ is the Fermi energy.
$\Theta$ is the step function that describes the available deexcitation channels.
In addition, the decay rate is double the energy width of quasi-particle state.\\

\begin{figure}
\center
\rotatebox{0} {\includegraphics[width=0.8\linewidth]{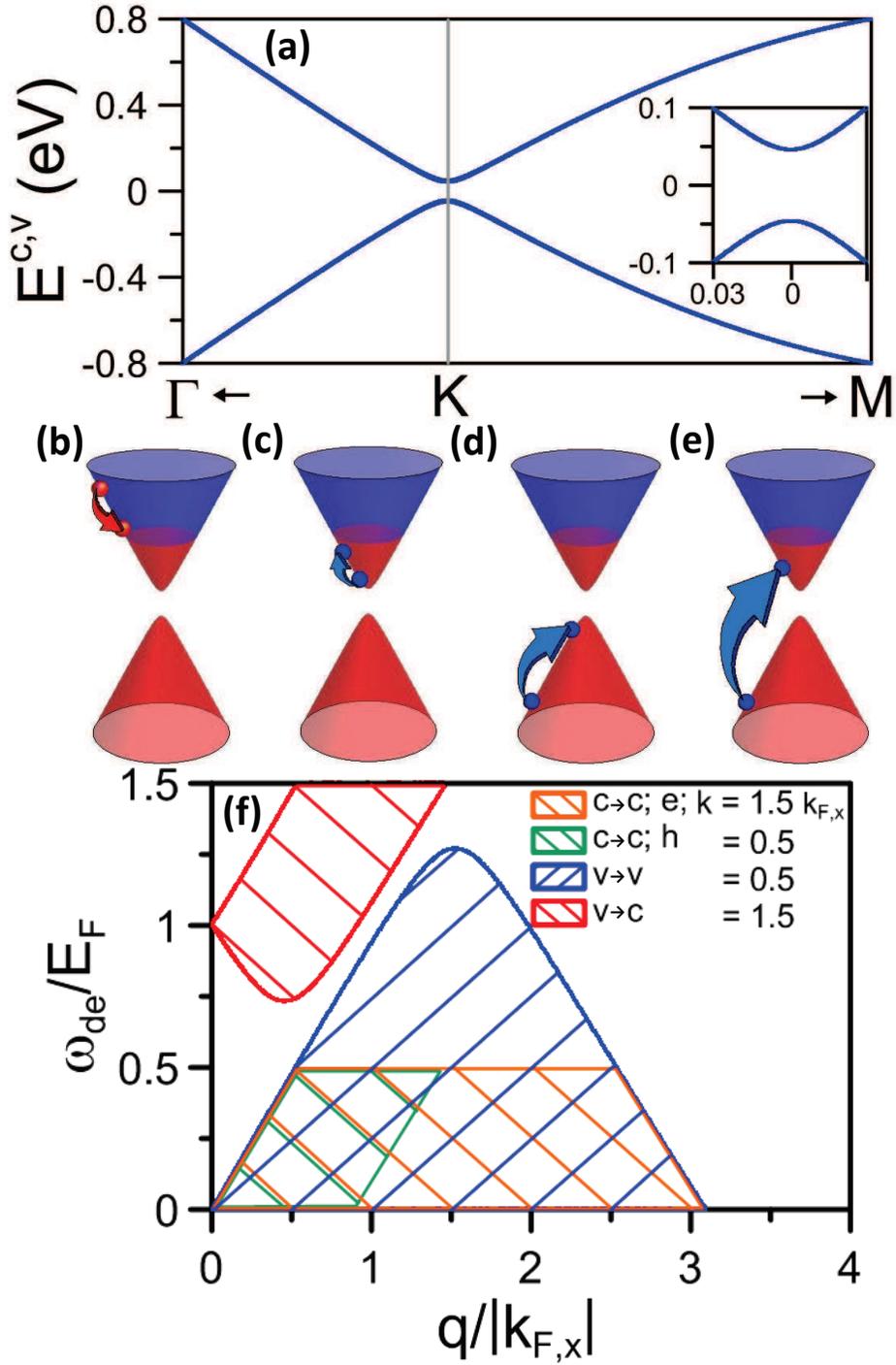}}
\caption{(a) The low-lying energy bands along the high-symmetry points, in which those near the Dirac points are shown in the inset. The available deexcitation channels of the specific excited states are indicated for (b) the conduction electrons, (c) the conduction holes,  and the valence holes scattered into the (d) same and (e) distinct bands. Furthermore, the relations between the deexcitation energies and transferred momenta are illustrated in (f).}
\label{Figure 2}
\end{figure}

Germanene displays the feature-rich band structure due to the significant SOC and the buckled honeycomb lattice.
The conduction band is symmetric to the valence one about the zero energy (Fig. 2(a)).
These two bands present parabolic energy dispersions near the K point, in which the separated Dirac points have an energy spacing of ${E_D=93}$ meV because of SOC (inset in Fig. 2(a)).
The state energy (${E^h(\bf k\,)}$) in this work is measured from the middle of energy spacing.
They gradually become the linear dispersions in the increase of the state energy.
The energy dispersions are anisotropic at sufficiently high energy ($|E^{c,v}|>0.2$), as observed along the K$\Gamma$ and KM directions.
With the increasing wave vector, the former exhibits more obvious changes, compared with the latter.
The anisotropic energy spectrum will play an important role in the Coulomb decay rates.\\

The Fermi energy/free carrier density dominates the main features of electronic excitations and thus determines the Coulomb decay channels.
When $E_F$ is in the middle of energy spacing, the excited electrons/holes can decay into conduction/valence band states by using the interband SPEs.
The increase of $E_F$ creates the intraband SPEs and plasmon mode, and induces the drastic changes in the interband SPEs.
Such Coulomb excitations will diversify the decay channels.
As for the excited conduction electrons, the final states during the Coulomb deexcitations only lie between the initial states and the Fermi momentum (a red arrow in Fig. 2(b)), according to the Pauli exclusion principle and the conservation of energy and momentum.
The available deexcitation channels, the intraband SPEs, make the most important contributions to the Colomunb decay rates for the low-lying conduction electrons, corresponding to the orange part in Fig. 2(f).
But when the initial state energy is high, the interband SPEs and the second/third kind of plasmon modes might become the effective deexcitation mechanisms (discussed later in Fig. 3).
Concerning the excited holes in the conduction band, they could be de-excited to the conduction states (c$\rightarrow$c; a blue arrow in Fig. 2(c)) through the intraband SPEs, owing to the low deexcitation energies and transferred momenta, as shown by the green part in Fig. 2(f).
On the other hand, the valence holes present two kinds of decay processes: v$\rightarrow$v and v$\rightarrow$c in Figs. 2(d) and 2(e), respectively.
Their available decay channels, respectively, cover [intraband SPEs, interband SPEs $\&$ the second/third kind of plasmon modes] and [interband SPEs $\&$ acoustic plasmon modes], corresponding to the blue and red parts in Fig. 2(f).
Specifically, the latter has the large deexcitation energies at small momenta and is thus expected to exhibit the efficient and unusual Coulomb decay rates.\\

\begin{figure}
\center
\rotatebox{0} {\includegraphics[width=0.8\linewidth]{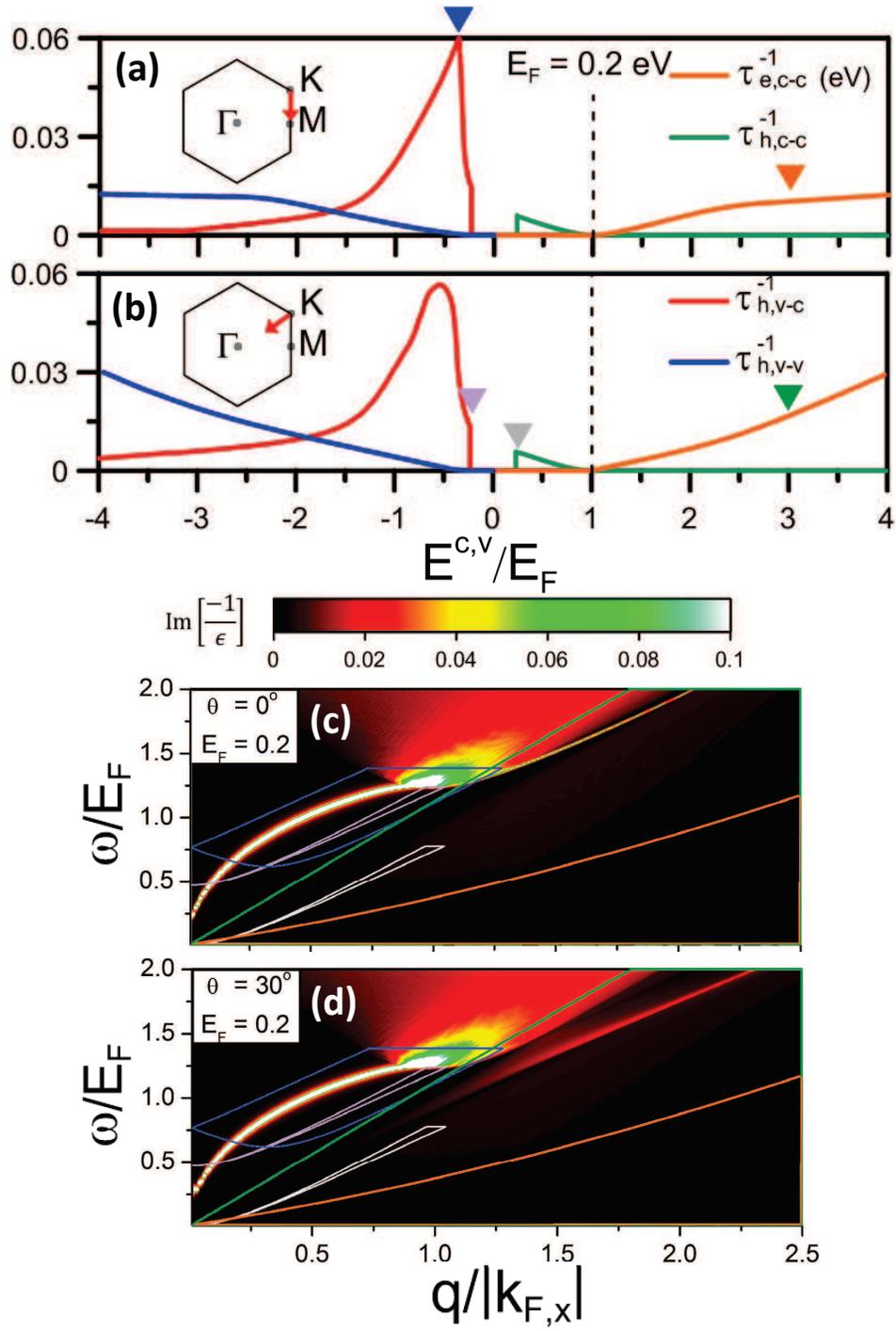}}
\caption{For monolayer germanene with ${E_F=0.2}$ eV, the Coulomb decay rates of the quasi-particle states along the (a) KM and (b) K$\Gamma$ directions, and the available deexcitation spectra due to the specific states indicated by arrows in (a) $\&$ (b) under: (c) ${\theta\,=0^\circ}$ and (d) ${30^\circ}$.}
\label{Figure 3}
\end{figure}

The Coulomb decay rates are very sensitive to the quasiparticle state ($\bf k$, $h$).
As to the excited conduction electrons, they have an infinite lifetime at the Fermi-momentum states, as shown in Figs. 3(a) and 3(b) by the orange curves.
${1/\tau_e}$ is vanishing at ${E^c=E_F}$ since all the electronic states are occupied below the Fermi level.
With the increasing conduction-state energy, the ${c\rightarrow\,c}$ intraband process is available.
The intraband SPEs make the main contributions to this process (the orange part in Fig. 2(f)), so that the decay rate monotonously grows with ${E^c}$.
When the excited states are close to $E_F$ ($|E^c-E_F|<$ 0.5 $E_F$), ${1/\tau_e}$ is roughly proportional to $(E^c-E_F)^2ln|E^c-E_F|$.
This energy dependence has been observed in 2D electron gas \cite{JEPT:33-997,PRB:26-4421}.
For the higher-energy conduction states, the Coulomb decay rates depend on the anisotropic energy bands.
Along the KM direction (Fig. 3(a)), ${1/\tau_e}$ increases and then reaches a saturated value after ${E^c>3E_F}$ (an orange arrow).
But for the K$\Gamma$ direction (Fig. 3(b)), it is getting large in the further increase of $E^c$ (a green arrow).
This important difference between the former and the latter lies in whether the second/third kind of plasmon mode and the interband SPEs are the effective deexcitation channels.
The higher-energy electronic states have the stronger energy dispersions along K$\Gamma$ (Fig. 2(a)), so that their deexcitation energies at large transferred momenta are consistent with those of the plasmon modes and interband SPEs.
For example, the conduction state of ${E^c=3E_F}$ along K$\Gamma$ has a lot of deexcitation channels indicated by the region below the green curve in Fig. 3(c) at ${\theta\,=0^\circ}$.
The similar results are revealed in the different momentum directions, e.g., the green curve at ${\theta\,=30^\circ}$ in Fig. 3(d).
The effective deexcitation channels cover the intraband SPEs, interband SPEs, and plasmon modes.
The latter two are responsible for the enhanced Coulomb decay rates in the high-energy conduction states along K$\Gamma$.
On the other hand, the ${E^c=3E_F}$ conduction electron along KM has the lower deexcitation energies and thus only exhibits the intraband SPEs, as shown by the regions below the orange curves in Figs. 3(c) and 3(d).\\

The deexcitation behaviors of the excited holes strongly depend on the conduction or valence states.
Concerning the conduction holes, the Coulomb decay rates are isotropic, as indicated by the almost identical $\tau^{-1}_{h,c-c}$'s along KM and K$\Gamma$ (green curves in Figs. 3(a) and 3(b)).
Furthermore, the energy dependence is similar to that of the low-lying conduction electrons (2D electron gas).
Such results directly reflect the fact that  the intraband SPEs are the only available deexcitation channels, e.g., the gray regions related to the conduction Dirac point (Figs. 3(c) and 3(d)).
Specifically, the K point (a gray arrow in Fig. 3(b)) has the largest Coulomb decay rate among all the excited conduction holes.\\

On the other side, the decay rates of the valence holes exhibit the unusual ${\bf k}$-dependences.
The valence Dirac point has a significant decay rate (a purple arrow in Fig. 3(b)), being much higher than that of the conduction one.
It only presents the ${v\rightarrow\,c}$ decay process, in which the deexcitation channels mainly come from the interband SPEs and the undamped plasmon modes, as indicated in the purple regions in Figs. 3(c) and 3(d).
They create the important difference between the valence and conduction Dirac points.
With the increase of the valence-state energy, two decay processes, ${v \rightarrow\,c}$ and ${v\rightarrow\,v}$, contribute to the Coulomb decay rates.
As to the former, the available range of the strong acoustic plasmon grows and then declines quickly for the low-lying valence holes, leading to an unusual peak structure in ${\tau^{-1}_{h,c-v}}$ at small $E^v$ (the red curve in Fig. 3(a)/Fig. 3(b)).
For example, the ${E^v=-0.4}$ $E_F$ valence state along KM has the widest plasmon-decay range enclosed by the blue curves in Figs. 3(c) $\&$ 3(d), so that it exhibits the fast Coulomb decay (a blue arrow in Fig. 3(a)).
The plasmon-induced deexcitations are almost absent for the deeper valence states (e.g., ${E^v<-1.5}$ $E_F$ along KM).
The interband SPEs also make some contributions to ${\tau^{-1}_{h,c-v}}$,  and they dominate the Coulomb decay rates of the deeper-energy states, e.g., the red curves along KM and K$\Gamma$ at ${E^v<-2 E_F}$.
Specifically, in the ${v\rightarrow\,v}$ process, the excited valence holes (the blue curves in Figs. 3(a) and 3(b) behave as the excited conduction electrons (the orange curves) in terms of the ${{\bf k}}$-dependence and the deexcitation channels.
The intraband SPEs are the dominating mechanisms in determining ${\tau^{-1}_{h,v-v}}$ of the low-lying valence states (${E^v>-2 E_F}$).
They are replaced by the intraband SPEs, the interband SPEs, and the second/third kind of plasmon mode for the deeper valence states along K$\Gamma$.
This accounts for the anisotropic Coulomb decay rates along K$\Gamma$ and KM.\\

\begin{figure}
\center
\rotatebox{0} {\includegraphics[width=1.0\linewidth]{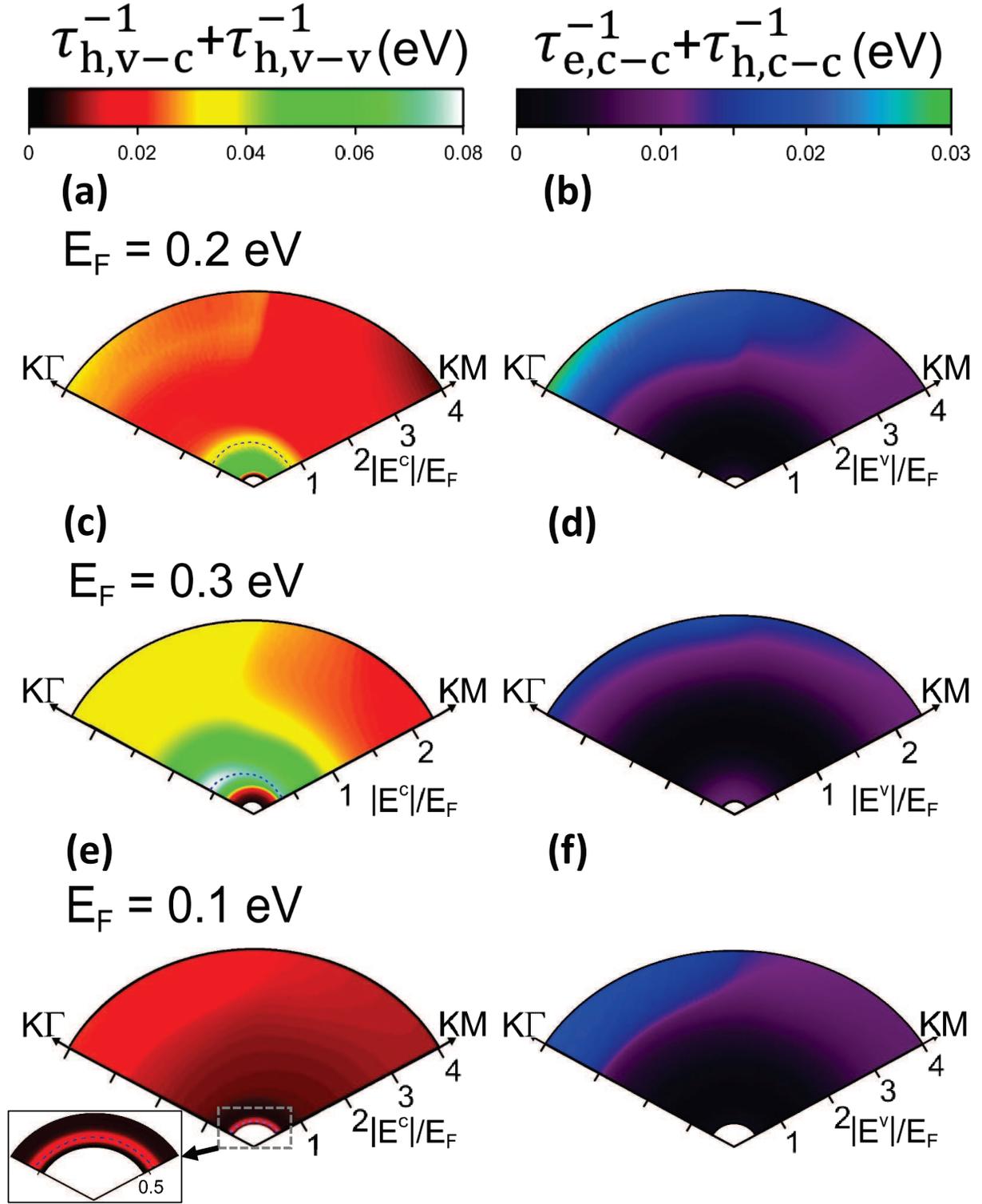}}
\caption{The wave-vector-dependent Coulomb scattering rates of (a) the valence holes and (b) the conduction holes and electrons at ${E_F=0.2}$ eV. Similar plots at ${E_F=0.3}$ eV and 0.1 eV, are respectively, shown in [(c) $\&$ (d)] and [(e) $\&$ (f)].}
\label{Figure 4}
\end{figure}

The wave-vector- and Fermi-energy-dependent Coulomb scattering rates, as clearly shown in Figs. 4(a)-4(f), deserve a closer examination.
The decay rates of the valence holes exhibit the oscillatory energy dependence along any specific direction of ${\bf k}$.
The strongest Coulomb scatterings, being associated with the undamped acoustic plasmons,  appear at valence states below the Dirac point (the dashed blue curves in Figs. 4(a), 4(c) and 4(e)).
The valence-state decay rates depend on the direction of ${\bf k}$, in which they are, respectively, lowest and highest along KM and K$\Gamma$.
Apparently, there exist the anisotropic Coulomb decay rates for any valence-state energies.
It is relatively easy to observe the anisotropic behavior at the higher Fermi energy, e.g., the decay rates of valence holes at ${E_F=0.3}$ eV (Fig. 4(c)).
This is closely related to the strong anisotropy of the deeper valence band.(Fig. 2(a)).
As for conduction holes and electrons, the Coulomb scattering rates, as measured from that of the Fermi-momentum state, present the monotonous energy dependences. The anisotropic deexcitations come to exist only for the higher-energy conduction states.\\

 In this work, the Coulomb scattering rates in monolayer germanene are investigated using the screened exchange energy, in which the excitation spectra are evaluated within the RPA.
The excited states cover the conduction electrons, conduction holes and valence holes, respectively, creating the decay processes: ${c\rightarrow\,c}$, ${c\rightarrow\,c}$, and [${v\rightarrow\,c}$ $\&$ ${v\rightarrow\,v}$].
The low-lying conduction electrons/holes present the isotropic scattering rates, mainly owing to the dominating intraband SPEs.
Furthermore, they behave as 2D electron gas in the energy-dependent Coulomb decay rates.
The other excited states exhibit the rich and unique ${\bf k}$-dependence, including the oscillatory energy dependence and the strong anisotropy.
Specifically, the low-energy valence states have the largest decay rates by means of the undamped acoustic plasmon modes, especially for that along KM.
Such deexcitation modes also lead to the important difference between the valence and conduction Dirac points.
The deeper valence states and the higher conduction states have the similar deexcitation channels, and so do the ${\bf k}$-dependent decay rates.
The intraband SPEs are replaced by intraband SPEs, interband SPEs and the second/third kind of plasmon mode as KM is transferred to K$\Gamma$.
This is responsible for the anisotropic decay rates.
It is relatively easy to observe the feature-rich Coulomb decay rates at higher Fermi energies.
The theoretical predictions on decay rates could be examined from the APRES measurements on the energy widths of quasi-particle states.

\newpage
\vskip 0.6 truecm
\par\noindent
\vskip 0.3 truecm
\renewcommand{\baselinestretch}{0.2}

\end{document}